\documentclass[a4paper]{article}
\usepackage{graphicx}
\usepackage{onecolpceurws}
\usepackage{booktabs}
\usepackage{wrapfig}

\def\copyrightspace{
  \long\def\@makefntext##1{\noindent ##1}
  \footnotesep 1em
  \footnotetext[0]{\em Copyright held by the author(s).}
  \footnotetext[0]{In: A.\ Garcia-Dominguez, G.\ Hinkel,  A.\ Boronat, and F.\ Krikava (eds.): Proceedings of the 13th Transformation Tool Contest, on-line, 17-07-2020, published at http://arxiv.org}
}

\title{Supporting Round-Trip Data Migration	 \\
for Web APIs: A Henshin Solution}

\author{
Daniel Str\"uber 
}

\institution{Radboud University Nijmegen \\ d.strueber@cs.ru.nl}

\begin{document}
\maketitle
\newcommand{\mone}{$M_1$-$M_2$-$M_1$}
\newcommand{\mtwo}{$M_2$-$M_1$-$M_2$}

\begin{abstract}
We present a solution to the Round-Trip Migration case of the Transformation Tool Contest 2020, based on the Henshin model transformation language.
The task is to support four scenarios of transformations between two versions of the same data metamodel, a problem inspired by the application scenario of Web API migration, where such a round-trip migration methodology might mitigate drawbacks of the conventional ``instant'' migration style.
Our solution relies on Henshin's visual syntax, which seems well-suited to capture the problem on an intuitive level, since the syntax is already similar to the scenario illustrations in the case description.
We discuss the five evaluation criteria \textit{expressiveness}, \textit{comprehensibility}, \textit{bidirectionality}, \textit{performance}, and \textit{reusability}.
\end{abstract}
\vskip 32pt

\section{Introduction.}

During the evolution of web-based systems, provided APIs may change over time.
Implementing API changes in an instant usually leads to severe complications, as developers of client software need to rapidly respond to such changes.
A desirable alternative is a \textit{round-trip migration system} that provides support for several API versions in parallel, while internally managing the data in such way that consistency is ensured.

To study how transformation tools may be an enabling technology for such a system, this problem is the subject of a case in the 2020 edition of TTC, called \textit{Round-Trip Migration of Object-Oriented Data Model Instances} \cite{Beurer2020}.
The case description includes four scenarios of changes that need to be supported by such a system: \textit{create/delete field}, \textit{rename field}, \textit{declare field optional/mandatory}, and \textit{multiple edits}.
The task is to develop transformations capturing these changes between two versions $M_1$, $M_2$ of the same metamodel.
In each scenario, four transformations are required: the migration and subsequent back-migration for both possible round-trip directions (\mone, \mtwo).

In this paper, we present a solution to the case based on the Henshin model transformation language \cite{Arendt2010,Strueber2017}.
Henshin is a model transformation language that supports the declarative, graph-based specification of in-place transformations.
The basic features of Henshin's tool set are a suite of editors and an interpreter kernel.
It also offers a variety of advanced features, such code generation for parallel graph pattern matching and support for various transformation analyses.


Henshin provides an expressive visual syntax that aims to support usability during transformation development \cite{Strueber2017}.
In fact, combining the object-diagram paradigm with change descriptions, Henshin's syntax is similar to the informal notation used in the case description.
Whereas the case description illustrates the change of specific instances, Henshin captures such changes in the form of reusable  \textit{rules}.
Rules express basic match-and-change patterns.
To specify control flow between rules where needed, Henshin provides \textit{composite units} that orchestrate the execution of a number of sub-units and rules.
From a variety of available units (including random execution and loops), our solution uses one:
Sequential units, supporting the specification of rules in a given order as well as the included data flow.

\section{Solution}

The solution comprises a set of Henshin rules and units, as well some Java-based glue code.
The  reason for having the glue code is to establish the connection  to the benchmark framework by implementing the provided task interface (constant over the different scenarios).
The solution contains 8 modules (one per scenario and case), encapsulating rules and units.
For scenarios 1--3, each module contains 2 rules, one for forward and one for background migration.
For scenarios 4, each module contains 2 units, each orchestrating 4 rules (one of them reused).
The full solution and setup instructions are available at https://github.com/dstrueber/ttc2020.

\subsection{Scenario 1: Create/delete field.}

\begin{figure}
	\centering
		\includegraphics[width=0.82\textwidth]{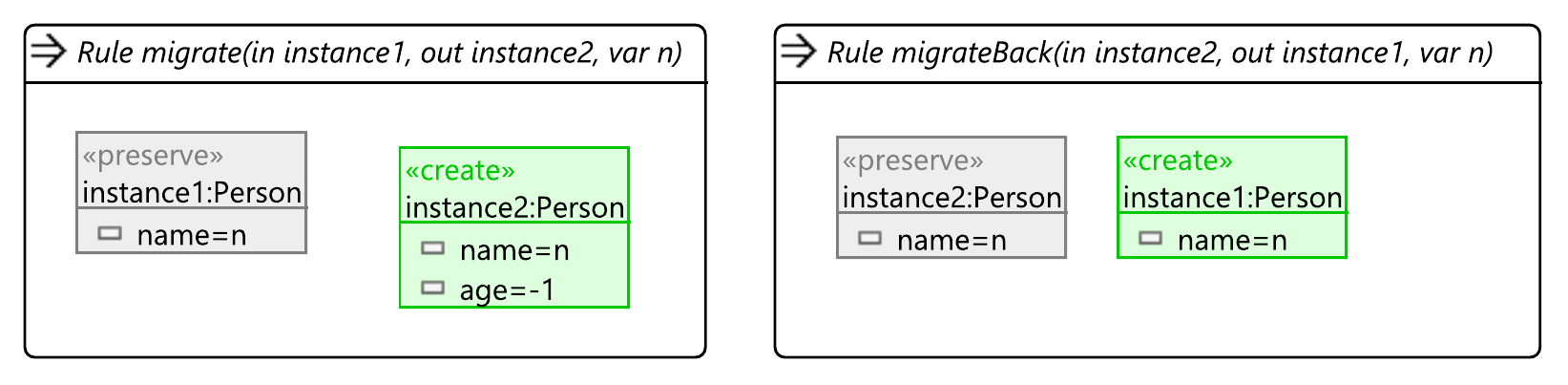}
		
		(a) Scenario 1, \mone\ case.
		\vspace{5pt}

		\includegraphics[width=1.00\textwidth]{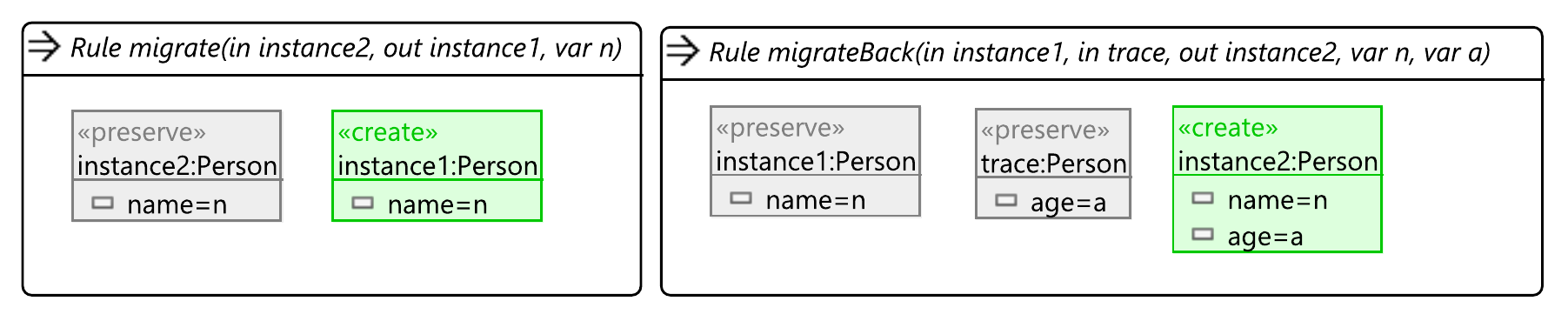}
		(b) Scenario 1, \mtwo\ case.
		
	\caption{Full solution for Scenario 1: create/delete field.}
	\label{fig:rules1}
\end{figure}

The solution for scenario 1 comprises four rules, shown in Fig.~\ref{fig:rules1}: two for both directions, one for migration and one for back-migration.
Each rule has a name and a parameter list; parameters have a name and one of the directions \textit{in}, \textit{out}, \textit{var}.
\textit{In} and \textit{out} parameters are passed into and out of the the rule from the usage context (in our case, the Java-based glue code), \textit{var} parameters (variables) are used internally to propagate information between different parts of the rule.
Each rule comprises a number of \textit{preserve} elements (shown in gray), specifying the context for creations, and \textit{create} elements (shown in green), specifying newly created elements.
For example, the top \textit{migrate} rule specifies a person from metamodel 1 as context for the creation of a new person based on metamodel 2, obtaining the same name and the age value of -1.
The \textit{migrateBack} rule for the \mtwo\ case requires a \textit{trace} person as a source for the age value.
The glue code for the previously applied \textit{migrate} rule ensures that this object is available and can be passed into \textit{migrateBack}. 

A usability limitation in this particular scenario (involving two metamodels that, from the user's perspective, are actually versions of the same metamodel) is that there is no visual distinction between elements from different metamodels.
This issue is mitigated by the use of names (e.g., \textit{instance1} and \textit{instance2}, referring to instances from $M_1$ and $M_2$) as well as the coloring of separate actions (preserve, create).

\subsection{Scenario 2: Rename field.}

\begin{figure}[t]
	\centering
		\includegraphics[width=1.0\textwidth]{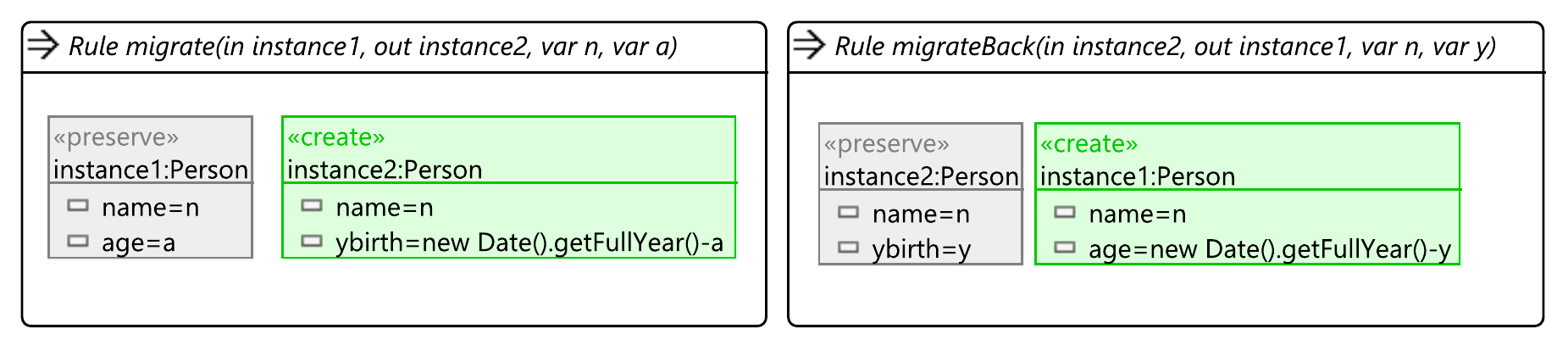}
\vspace{-20pt}
	\caption{Scenario 2: rename field (\mone\ case. The rules for \mtwo\ look the same).}
	\label{fig:rules2}
\end{figure}

Figure \ref{fig:rules2} shows two out of the four rules for scenario 2, capturing the \mone\ case (for symmetry reasons, the other two rules are identical).
In attribute calculations, Henshin supports the use of JavaScript expressions, including standard library objects such as \textit{Date}, as we use to specify the conversion between birth dates and age.
During execution, JavaScript expressions are evaluated using the Nashorn JavaScript engine.

\begin{figure}[t]
	\centering
		\includegraphics[width=1.0\textwidth]{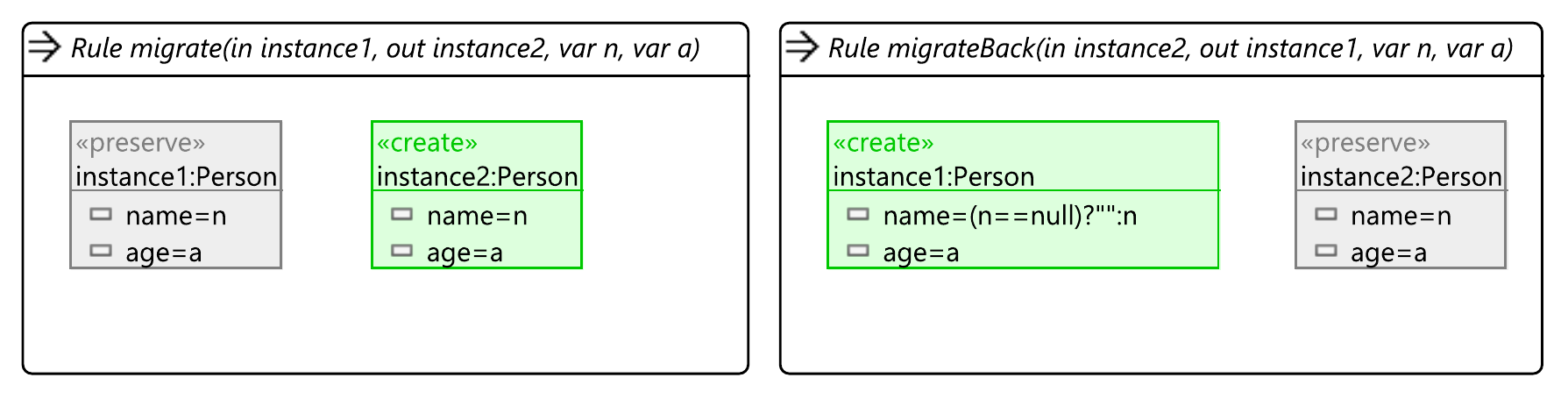}		
\vspace{-20pt}
	\caption{Scenario 3: Declare field optional/mandatory (\mone\ case. The rules for \mtwo\ look similar).}
	\label{fig:rules3}
\vspace{+10pt}
\end{figure}

\subsection{Scenario 3: Declare field optional/mandatory.}
Figure \ref{fig:rules3} shows two out of four rules for scenario 3, capturing the \mone\ case.
Again, we rely on Henshin's capability of using JavaScript calculations in attribute values.
In the \textit{migrateBack} rule, we use the ternary operator to check if the name of the provided person object is null, and specify different calculation outcomes for both cases.
The \mtwo\ case (not shown) is largely similar; the main difference is that the \textit{migrate} rule includes a handling of empty strings, in the same way that the shown \textit{migrate}rules handles the \textit{null} case.

\subsection{Scenario 4: Multiple edits.}
While scenarios 1--3 are straightforward, scenario~4 has a design space of possible solutions.
The rationale of our solution is to illustrate the reuse capabilities of Henshin, as requested in the case description.
Still, we notice an inherent trade-off between reuse, simplicity and performance.
The granularity of reuse in this case is so fine that the price of reuse appears high:
Compared to a solution without reuse  (capturing the edits in a single rule), the specification is much larger and therefore harder to understand and less efficient to execute.

\begin{figure}
	\centering
		\includegraphics[width=0.9\textwidth]{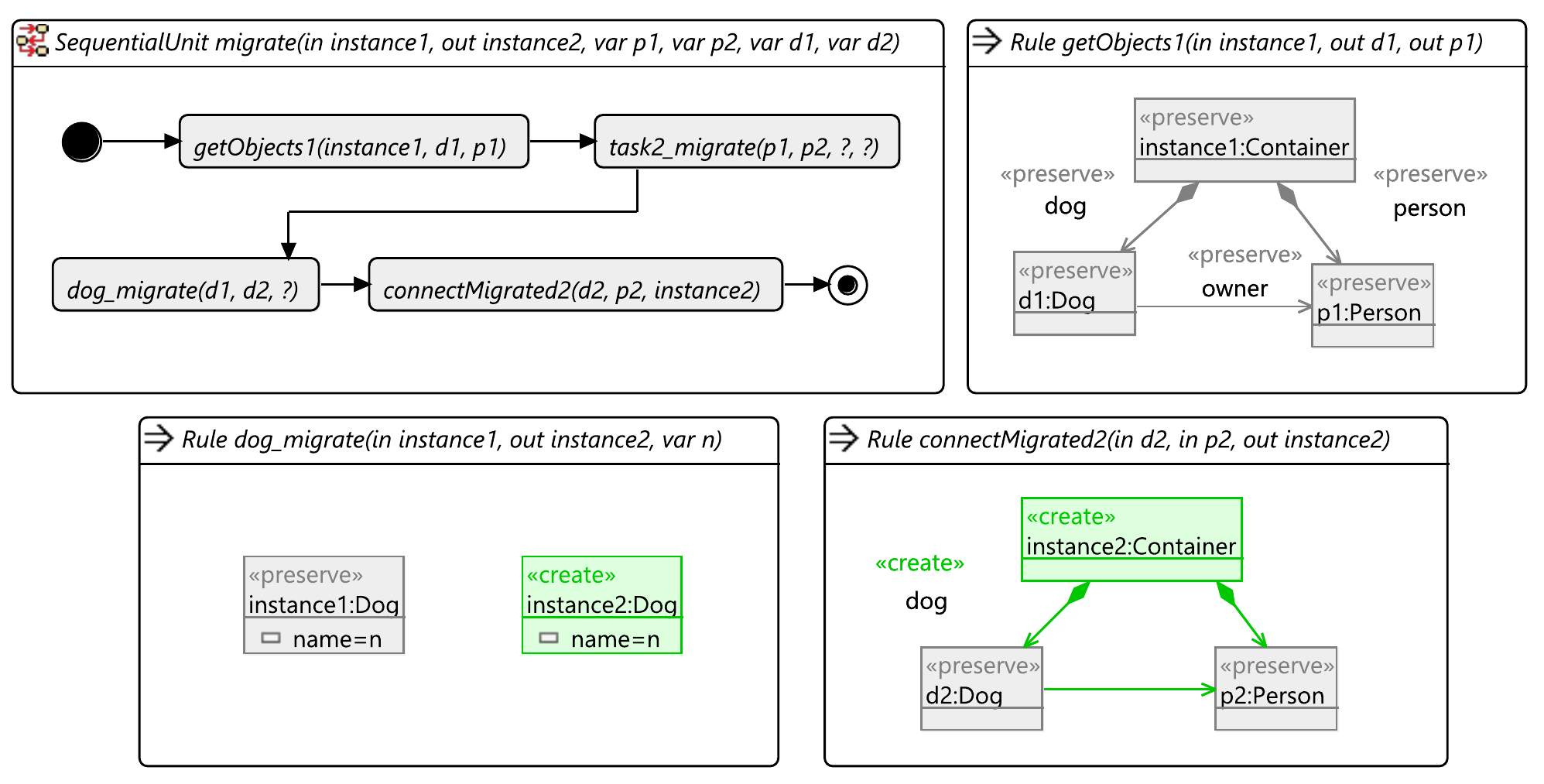}
		
	\caption{Scenario 4: Multiple edits; M1-M2-M1 \textit{migrate} step (\textit{migrate-back} and M2-M1-M2 look similar).}
	\vspace{15pt}
	\label{fig:rules4}
\end{figure}

Figure~\ref{fig:rules4} shows the implementation of the \textit{migrate} step in the \mone\ direction, comprising a sequential unit that invokes four rules -- three of them presented with the unit, one of them reused from Scenario 2.
(For reading convenience, the rule from Scenario 2 was actually copied and included into the Scenario 4 transformation file, but it would be possible to maintain an actual reference as well---using EMF's concept of remote references.)
The unit produces an instance of a $M_2$ container object from an instance of a $M_1$ container object, and also has four internal \textit{var} parameters.
The order of the rule invocations is specified in an activity-diagram-like notation.
Data flow is specified by passing parameters between rule invocations (questions marks refer to variables, which do not have to be set from the context).
Rule \textit{getObject1} fetches the person and dog objects from the container.
Rules \textit{task2\_migrate} and \textit{dog\_migrate} produce the migrated objects.
Rule  \textit{connectMigrated2} connects the migrated objects with a reference and encapsulates them in a container, which is yielded as~output.

The solution for \textit{migrateBack} and both steps in the \mtwo\ direction look largely similar. 
A small additional sophistication is that \textit{migrateBack} for \mone\ requires a copy (a.k.a. trace) of the previously migrated dog instance, which we implement in a similar manner as in the \mtwo\ case of scenario 1.

\section{Evaluation}

We discuss our solution in the light of the five evaluation criteria from the case description \cite{Beurer2020}.

\medskip

\noindent{}\textbf{Expressiveness.}
The assessment of expressiveness is based on the number of passing test cases: the more test cases a solution covers, the more expressive it is.
According to this criterion, the present solution is maximally expressive, since it passes all test cases.

\medskip

\noindent{}\textbf{Comprehensibility.}
With our solution, we aimed at providing a primarily declarative solution.
We achieved this goal by specifying all parts of the change logic using Henshin's declarative rule and control flow concepts.
In addition, our solution includes some glue code, written in Java, to implement the provided \textit{Task} interface.

\medskip
\noindent{}\textbf{Bidirectionality.}
Our solution (like any solution passing all test cases) is inherently bidirectional, since it supports transformations in both directions.
Still, Henshin does not provide dedicated support for automatically generating the transformation in either direction, and hence, leaves the effort for bidirectionality implementation to the user.
An extension of Henshin with dedicated support for bidirectionality exists in an early stage \cite{ermel2012visual}.

\begin{wrapfigure}{r}{0.55\textwidth} 
  \begin{center}
    \includegraphics[width=0.53\textwidth]{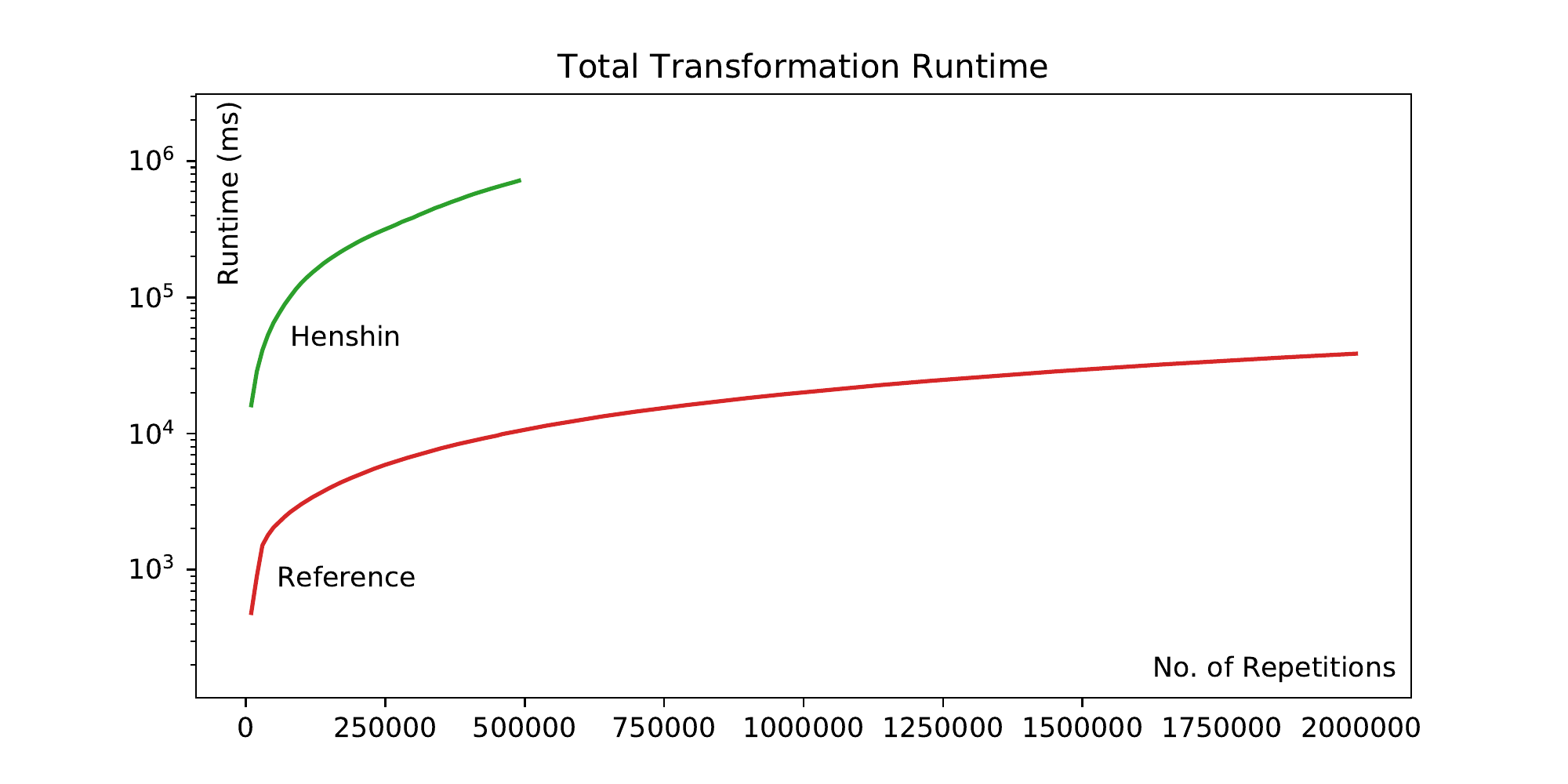}
		\vspace{-20pt}
  \end{center}
  \caption{Runtime measurements}
	\label{fig:expdata}
\end{wrapfigure}

\medskip
\noindent{}\textbf{Performance.}
Figure~\ref{fig:expdata} shows the execution time data from running the performance test on a dual-core i7-6600U CPU Ubuntu Linux system with 16 GB RAM.
We observe a large slowdown, by two orders of magnitude, compared to the Java reference implementation.
This slowdown is to be expected due to the overhead of Henshin's generic matching engine, and the decision to use a deliberately inefficient implementation (in favor of reuse) in the most complicated scenario (scenario 4).

\medskip
\noindent{}\textbf{Re-usability.}
Henshin's composite units support the specification of a control flow.
In scenario 4, we use this feature to reuse rules originally specified for the implementation of scenario 2.
Doing so, we enable reuse by following a suggestion from the case description: \textit{"delegating the migration of the referenced Person instances to migration functions which have been already defined."} \cite{Beurer2020}
Henshin has another reuse concept for specifying rule variants \cite{StruberRACTP18}, but this concept was not applicable to the given case, since it does not support variations on the level of node types (dog vs. person). This experience emphasizes the need for more expressive reuse concepts for model transformations \cite{chechik2016perspectives}.

Furthermore, it can be emphasized that the Java-based glue code for implementing the task interface is constant over the different scenarios.
Even though scenario 4 is the most complex scenario by far, we were able to fully reuse the glue code developed for scenarios 1--3 without further modifications.

\section{Outlook}

The trade-off related to reusability vs. simplicity/performance motivates further work on composition of transformation systems:
The complicated specification of units and rules in scenario 4 could be translated into a single rule, which might both increase readability and execution performance.
While there are various works on composition of rules from smaller parts \cite{Rensink10,AnjorinSLS14,fritsche2018short}, no existing work seems to consider control flow in the input transformation specification.
The considered round-trip migration scenario is also relevant during the model-driven migration between different backend versions of content management systems \cite{PrieferKS17}.

\medskip
\noindent{}\textbf{Acknowledgement.}
Thanks to Antonio Garcia-Dominguez for providing an initial version of the performance comparison figure.

\bibliographystyle{alpha} 
\bibliography{bib}

\end{document}